\documentclass[iop]{emulateapj}

\usepackage{verbatim}

\usepackage[]{natbib}
\usepackage{color}

\shorttitle{NIR spectroscopy of HD~100546}
\shortauthors{Brittain et al.}

\begin{document}

\title{NIR spectroscopy of the HAeBe star HD~100546: III. Further Evidence of an Orbiting Companion?}

\author{Sean D. Brittain}
\affil{Department of Physics \& Astronomy, 118 Kinard Laboratory, Clemson University, Clemson, SC 29634, USA }
\email{sbritt@clemson.edu}

\author{John S. Carr}
\affil{Naval Research Laboratory, Code 7211, Washington, DC 20375, USA}

\author{Joan R. Najita\altaffilmark{1}}
\affiliation{National Optical Astronomy Observatory, 950 N. Cherry Ave., Tucson, AZ 85719, USA}

\author{Sascha P. Quanz \& Michael R. Meyer}
\affiliation{ETH Zurich, Institute for Astronomy, Wolfgang-Pauli-Strasse 27, 8093 Zurich, Switzerland}

\altaffiltext{1}{Institute for Theory and Computation, Harvard-Smithsonian Center for Astrophysics, 60 Garden Street, Cambridge, MA 02138, USA}

\begin{abstract}
We report high resolution NIR spectroscopy of CO and OH emission from the 
Herbig Be star HD~100546. We discuss how our results bear striking resemblance to several theoretically 
predicted signposts of giant planet formation.
The properties of the CO and OH emission lines are consistent 
with our earlier interpretation that these diagnostics provide indirect evidence for a 
companion that orbits the star close to the disk wall (at 
$\sim$13~AU). The asymmetry 
of the OH spectral line profiles and their lack of time variability are consistent with emission 
from gas in an eccentric orbit at the disk wall that is approximately stationary in the inertial frame. 
The time variable spectroastrometric properties of the CO 
v=1-0 emission line point to an orbiting source of CO emission with an 
emitting area similar to that expected for a circumplanetary disk ($\sim$0.1~AU$^2$) assuming the
CO emission is optically thick.  We also consider a 
counterhypothesis to this interpretation, namely that the variable CO emission arises from 
a bright spot on the disk wall. We conclude with a brief suggestion of further work that can 
distinguish between these scenarios.

\end{abstract}


\section{INTRODUCTION}
\label{sec:intro}

Planet formation is a complex and poorly understood process \citep{Helled2013}.  
Yet candidate forming planets have been inferred in several protoplanetary disk systems 
based on studies of their spectral energy distributions and through direct imaging 
in the continuum (e.g. \citealt{Quanz13,Kraus12,Hue11}). A low mass accreting object
in a transition disk has also been imaged in H$\alpha$ \citep{Close14}.
The Herbig B9e star HD~100546 has attracted recent attention as a system that may 
have forming low-mass, possibly planetary, companions (e.g. \citealt{Quanz13,Mulders13,Britt13,Lisk12,Panic14}). 
Located at a distance of 97$\pm$4\,pc \citet{vL2007}
and with an age of $\sim$3-10~Myr \citep{Manoj06,vdA97}, 
HD~100546 has a spectral energy distribution (SED) with 
a minimal near infrared (NIR) excess coupled with 
a substantial mid/far infrared (M/FIR) excess, 
i.e., it is a transition disk object.  
Modeling the SED, \citet{Bouwman03} concluded that 
HD~100546 has a small disk in the inner few AU and a much more massive disk 
beyond $\sim$10\,AU  (see also 
\citealt{Liu03, Grady05, AvdA06a, Benisty10, Quanz11a, Panic14}). 
Millimeter observations indicate that the dust mass of 
the disk is  $5\times10^{-4}$M$_{\Sun}$ \citep{Henning98}. 
The large optically thin region between the inner disk and the outer 
disk suggests that the disk structure 
has evolved, possibly as a result of planet formation. 

Stellar accretion is also ongoing in the system. 
Balmer lines with redshifted absorption components at $\sim$220 km s$^{-1}$, 
consistent with magnetospheric accretion,  
have been observed toward HD~100546 \citep{Gui06}. 
Modeling the Balmer discontinuity of HD~100546, 
Pogodin et al.\ (2012) concluded that the accretion rate is 
$\rm 6\times10^{-8}~M_{\Sun}~yr^{-1}$. 
Despite this high accretion rate, 
the optically thin region of the disk appears to be clear 
of molecular gas.  
High resolution infrared spectroscopy of ro-vibrational H$_2$ 
\citep{Carmona11}, OH (\citealt{Lisk12}; Paper I), and
 thermal-fluorescent CO emission \citep{Britt09, vdP09} 
find no molecular gas within $\sim$13\,AU of the star. Similarly, 
\citet{Habart2006} find that this inner region is also devoid of PAH emission. 
Some atomic gas remains in the inner region of the disk, 
based on [\ion{O}{1}] 6300\AA\ spectroscopy 
\citep{AvdA06a},  which is not surprising given the continued stellar accretion. 
Thus SED modeling, interferometry, direct imagery, and spectroscopy
of warm molecules are all consistent with HD~100546 being an accreting transitional
disk system whose inner 13~AU has been mostly (though not entirely) emptied
of material.

Careful analysis of the CO and OH high resolution spectroscopic 
observations provides possible evidence of an orbiting object 
near the disk wall at 13\,AU.  
The asymmetric line profiles that we observed in the OH spectrum of
HD~100546 are consistent with emission from an annulus with
an eccentricity of $\sim$0.2 (Paper I). 
Such a level of eccentricity can be produced by tidal interactions
with a high mass planetary companion on a circular orbit \citep{KD06}. Because
dynamical models predict that the semi-major axis of the eccentric
annulus will precess very slowly ($\sim$10$\degr$/1000 orbits; \citealt{KD06}), 
we noted that if the
asymmetric OH emission from HD~100546 arises from an eccentric
annulus induced by a planetary companion, the line profile should 
not vary over timescales of months to years (Paper I).

The presence of an orbiting companion has also been
inferred from non-axisymmetric structure in the
gaseous CO v=1-0 emission from HD~100546 (\citealt{Britt13}; Paper II). 
In contrast to the CO hotband emission from HD~100546, which showed 
no variability from 2003 to 2010, 
the line profile and spectroastrometric 
signal in the v=1-0 CO emission from HD~100546 
showed significant variability over the same period. 
We showed that the CO hotband emission could be explained as emission 
arising from an eccentric inner rim and 
an (axisymmetric) circumstellar disk beyond 13\,AU. The spectroastrometric 
measurement of these emission lines confirmed the sense of rotation of the 
disk inferred from the spectroastrometric analysis of the [\ion{O}{1}] 6300\AA\ line 
\citep{AvdA06a}. The CO v=1-0 emission could be fit as emission from 
a similar non-time-variable component 
(an eccentric inner rim and 
an axisymmetric circumstellar disk beyond 13\,AU) 
plus a compact source of excess emission that varies
in position and velocity as it orbits the star.  
The lower limit on the 
emitting area ($\sim$0.1AU$^2$) of the orbiting component is 
similar to, but smaller than, the size predicted for a circumplanetary 
disk around a 5~M$\rm_{Jup}$ planet at a distance of 13\,AU from HD~100546 
\citep{AB09a, ML11a}.  

Thus, previous work offers evidence for a companion
forming near the disk wall of HD~100546 and
suggests ready made observational tests of this interpretation.
In Paper II we noted that if this interpretation is 
correct the excess CO v=1-0 emission would become blue-shifted after 2010
while the line profiles of the hot band CO transitions would remain unchanged. 
Furthermore, the spectroastrometric signal of the blue side of the v=1-0 line
would become offset toward the east. In this paper
we present results from subsequent high resolution
$L-$ and $M-$ band spectroscopy of HD~100546 in 2013
which confirm our interpretation of the previous observations
of ro-vibrational OH and CO emission from the HD~100546 disk. 
Observations such as these can potentially help us to understand 
the impact of forming planets on the circumstellar disk.  They may also 
bear on the existence and nature of circumplanetary disks 
and the nature of the planetary accretion process. 

\section{Observations and Data Reduction}
\label{sec:Observations}
Near-infrared spectra of HD~100546 were acquired with CRIRES on the \emph{European Southern 
Observatory Very Large Telescope (VLT)} on March 18, 2013.
The data were acquired with the 4 pixel (0$\arcsec$.4) slit without the use of adaptive optics. 
This instrumental configuration provides a resolving power of R$\sim$50,000.

The $L-$band observations were taken with the slit at a position angle (PA) of 
90$\degr$ east of north. The $M-$band observations were taken with the slit at 
a PA of 90$\degr$ and 140$\degr$ as well as at the corresponding anti-parallel positions. 
In this paper we restrict our analysis to direct comparison to the previous results during 
which the slit PA was 90$\degr$. In a subsequent paper 
we will present the full analysis and modeling of the full data set. 

The full width at half maximum of the point spread function (PSF) of the 
continuum ranged from $0.\arcsec$5 - $0.\arcsec$7 
and the airmass of the observations ranged from 1.3-1.6. A summary of 
observations is presented in Table \ref{tab:observations}.
Observations in the $L$- and $M$-band are dominated by a strong thermal background. 
Therefore, an ABBA nod pattern
 between two positions separated by $\sim 10\arcsec$ was used to cancel the thermal background to 
 first order.  The CO emission can be detected out to $\sim$0.7$\arcsec$ in each direction 
 \citep{Britt09}, thus there is 
 no contamination between the beams. 

The data are reduced using a customized set of routines based on algorithms developed 
for the reduction of data acquired with PHOENIX and NIRSPEC 
(see \citealt{Britt07a}). Each AB pair of observations is 
 combined in the sequence ($A-B$) and then divided by the normalized flat field image. 
 The nod positions were jittered in a 2$\arcsec$ box, so the median of the aligned images could be 
 used to identify cosmic rays and bad pixels. 
 The  median of the aligned images was then compared to each of the individual frames. Pixels 
 differing by more than  $6\sigma$ from the median were rejected in the average. 
  
The centroid of the PSF was measured by fitting a Gaussian curve to each column. 
For the $M-$band spectra, the centroid measurements 
 of the parallel and anti-parallel beams 
 were averaged to find the spectroastrometric offset of the spectra
relative to the continuum. 
 The 1-D spectra are formed by summing the 15 rows (1.3$\arcsec$) 
 centered on the PSF.  The spectra were then wavelength calibrated
 by fitting an atmospheric transmittance model generated by the Spectral Synthesis Program \citep
 {1974JQSRT..14..803K}, 
which accessed the 2000HITRAN molecular database \citep{2003JQSRT..82....5R}.  
 The 1-D spectrum of HD~100546 was divided by the spectrum of the hot star standard $\alpha$Cru.
 The telluric standard was observed at a similar airmass and scaled to provide an exact match 
 (Figures  1 and 2). The S/N of the spectra 
 and the fidelity of our centroid measurements are summarized in Table 1.

\section{Results}
\label{sec:Results}
\subsection{OH emission}
We detect the P4.5 (1$\pm$), 8.5(2$\pm$), 9.5(2$\pm$), and 10.5(1$\pm$) lines at 1-4\% of the 
continuum 
(Fig. 1). The equivalent widths of these 
lines were converted to relative fluxes using a spectrophotometric measurement of HD~100546 
taken with 
the \emph{International Space Observatory}. While we found that HD~100546 is 
variable in the NIR (Paper~II), we assume for this analysis that the shape of the spectral 
energy distribution in 
the $L-$band has not varied. The 
relative flux of each line is presented in Table 2. By performing a linear least square fit to the points 
in the excitation diagram, we find that the 
rotational temperature of the OH gas is 1060$\pm$40~K --
consistent with the upper limit of 1400K inferred from the previous observation. 
The equivalent width of the OH emission 
lines decreased by 50\% from December 2010 to March 2013. This can be accounted
for by a brightening of about 1magnitude in the $L-$band -- consistent with the photometric 
variability of this object in the NIR (Paper II); though we cannot rule out variation in 
the line fluxes.

We compare the profiles of the OH observed in 2013 with CRIRES on the VLT and 
2010 with PHOENIX on Gemini South (Fig. 3a). In Paper I we
showed that the asymmetric line shape could
be fit if most of the emission arises from an eccentric annulus near disk wall.
  Over the span of 28 months, we see 
no evidence of variation of the line shape\footnote{The use of the 0.4" slit 
is important in this respect. If we had observed the disk with the 0.2" slit with 
AO, then the observed line shape would depend sensitively on the pointing 
and the portion of the disk wall sampled\citep{HB13}.} as predicted in Paper I 
if the eccentricity inferred from the 
line profile is induced by a companion.

\subsection{CO emission}
We detect numerous fundamental ro-vibrational CO lines. For this paper we 
narrow our focus on the transitions overlapping 
our previous observations near the v=1-0 P26 line 
(2029~cm$^{-1}$ to 2035~cm$^{-1}$; Fig. 2). Comparison of the hot band lines observed in 2013 to 
the previous epochs shows only subtle variability in the shape of these lines  (Fig. 3b). 
However, like the OH lines, the equivalent widths of the 
hot band lines have varied. The 2013 equivalent widths are lower than those in 2010 
by 30\% and are consistent with the values measured in 2003. 
Despite these variations, the spectroastrometric signal of the hotband lines is 
similar in all four epochs and is consistent with 
emission from the circumstellar disk.

 In contrast to the hot band lines, the line profile and 
 spectroastrometric signal of 
the v=1-0 P26 line have continued to vary relative to previous epochs (Figs. 3b, 4a-4d).
We analyze the CO v=1-0 emission 
using the procedure and rationale outlined in Paper II. 
We first normalize the spectrum so that the CO hotband lines have the 
same equivalent width as in the 2003 spectrum. 
Table 3 shows the scaled EW of the CO v=1-0 emission in the resulting 
spectrum. 
We then subtract the 2003 spectrum to obtain the spectrum of the 
CO v=1-0 excess emission component (Fig. 4). 
The EW of the excess CO emission component and its 
velocity centroid and FWHM are shown in Table 3 where 
they are compared with the values from all earlier epochs.  
 
As described in Paper II and summarized in Table 3, 
between 2003 and 2006, the red side of the P26 line brightened, 
the spatial offset of the red side of the line decreased, 
and the CO excess emission component had 
a velocity centroid of 
$+6\pm1$km~s$^{-1}$  
(compare Fig. 4a and Fig. 4b). 
Similarly, in 2010 the P26 line brightened further, 
and the excess emission was centered near zero velocity ($-1\pm1$km~s$^{-1}$).
In this part of the line profile, the spatial centroid of the line became 
offset further to the east (Fig. 4c). 
In the 2013 epoch reported here, the blue side of the P26 line 
is again brighter than in 2003 and the excess is now
blue shifted ($-6\pm1$km~s$^{-1}$). The spatial centroid 
of the red side of the line is comparable to that in 2003 (Table 3), 
but the blue side of the line is now extended further to the east (Fig. 4d). 

 \section{Discussion}
 \label{sec:Discussion}
\subsection{Orbital Analysis}
In Paper II, we suggested that the variations in the v=1-0 line 
emission could be explained by the presence of a 
spatially concentrated source of CO emission that orbits the star within the disk wall. 
A schematic of this scenario is shown in Figure 4e. 
We can obtain a rough constraint on the orbit of the CO excess component 
given the velocity centroids observed in 2006, 2010, and 2013. 
Assuming a system inclination of 42$\degr$ \citep{Ardila07, Pineda14} and 
a stellar mass of 2.4~M$_{\sun}$, we fit a circular orbit 
to the measured velocities with the 
orbital radius $R$ and the orbital phase of the excess in 2003 
as free parameters. The result of a $\chi^2$ fit (Fig. 5) gives $R = 12.9^{+1.5}_{-1.3}$~AU
and an orbital phase of $\phi = 6\degr^{+15\degr}_{-20\degr}$, where
$\phi = 0\degr$ corresponds to the NW end of the semimajor axis. If we adopt a 
higher inclination (e.g. 50$\degr$; compare for example \citealt{Quanz11a} and \citealt{Panic14}), 
our best fit shifts to $R = 14.0^{+1.6}_{-1.3}$~AU
and an orbital phase of $\phi = 15\degr^{+13\degr}_{-15\degr}$

Hence, the velocity centroids at the three measured epochs 
are consistent with circular motion and 
locate the excess source in the vicinity of the 
disk wall. For an orbit with $R\simeq 12.5$\,AU, 
the inferred orbital phase would locate the excess emission behind 
the disk wall (in projection) in 2003, hiding it from view, consistent with 
our use of 2003 as the reference epoch in this analysis ( see also Paper II and Mulders et al. 2011). 
Figure 4e shows the velocities and implied position angles of the excess 
component. 

To demonstrate that our spectroastrometric results
are consistent with the above picture, 
we modeled the spectroastrometric signal 
that would be produced by the disk plus an
extra source of CO emission in a circular orbit, 
following the method described in Paper~II. 
Briefly, a CO emission component with the observed line profile of the excess emission 
is placed at the appropriate radius and phase in its orbit and 
added to our earlier model of CO emission from an eccentric disk wall
and an axisymmetric circumstellar disk. Synthetic spectroastrometric signals 
are then calculated and compared with the observed values (Fig.\ 4a-4d). 

In our model, the 5$\micron$ continuum arises from a compact 
source of emission within 1\,AU of the star \citep{Panic14}, thus to a very good approximation
the center of the PSF is aligned with the star (within $\sim$1\,mas). 
The slit was aligned E-W in all four epochs presented
in this paper, so the center of the PSF falls on the N-S line in our schematic (Fig. 4e). 
For simplicity we adopt R=12.5~AU and phase=0\degr in 2003 for our calculation. 
As a result,
the spectroastrometric signal of the P26 line 
at this epoch is consistent with emission arising from an 
axisymmetric  disk  (Fig. 4a).
As shown in Figure 4b-4d, the observed spectroastrometric signals in 
the subsequent epochs are well fit by our model
of a compact source of emission in a Keplerian orbit near the disk wall, supporting our earlier interpretation (Paper II). 


\subsection{Comparison with Previous Results}
In Papers I \& II, we discussed two new lines of evidence for a massive substellar companion
within the disk gap of HD 100546.
In the first paper, we noted that the 
ro-vibrational OH emission lines from the disk are 
asymmetric and that the observed line profile can be 
explained by a disk with an eccentric inner rim. 
The OH line profiles 
can be reproduced if 75\%  of the OH emission
arises from an annulus with an  eccentricity of 0.18$\pm 0.11$
while  the remainder arises from a circular outer disk.
An inner rim with this level of eccentricity  
can arise from tidal interactions with a high mass companion (M$\geq \rm 3M_{Jup}$; e.g., \citealt{KD06}).
The new observations reported here provide important 
support for the interpretation presented in Paper I.
Dynamical studies that show how a giant planet
can induce an eccentric inner rim also
predict that the semi-major axis of the eccentric
rim will precess very slowly ($\sim$10$\degr$/1000 orbits; \citealt{KD06}). 
Hence, the asymmetric line profile from the rim is expected to be
approximately constant in time. 
The OH line profiles we observe are consistent with this picture (Fig. 3a). 

In the second paper, 
we presented multi-epoch rovibrational CO emission and showed that the spectral 
shape and spectroastrometric signal of the v=1-0 lines vary while 
those of the hotband lines do not.
We showed that the variability of both the spectral shape and 
spectroastrometric signal of the v=1-0 lines could be explained if a 
compact source of v=1-0 emission orbits the star near the inner rim of the 
outer disk (R$\sim$13~AU). 
We further suggested that the source of this orbiting CO emission
is a circumplanetary disk. 
In the new CO observations reported here, both the velocity 
of the excess CO emission and the spectro-astrometric signal reported here
confirm our earlier interpretation of an emission source on
a circular Keplerian orbit near the inner wall of the disk.

\subsection{Origin of CO emission: Circumplanetary Disk or Companion/Wall Interaction?}
As discussed in Paper II, the luminosity of the 
excess CO emission is consistent with optically thick emission from a 
region with an emitting area of $\sim 0.1 - 0.2$\,AU$^2$ for
a temperature of 1400\,K and a 3 km~s$^{-1}$ intrinsic line width. 
This emitting area is a fraction of
the size of a circumplanetary disk that is expected for a $5\,M_J$
planet located 13 AU from a 2.4 M$_\sun$ star.  
The assumed temperature of 1400\,K seems plausible, because
it is similar to that found for the inner circumplanetary disk
in 3D radiation hydrodynamical simulations \citep{KK06, Gressel13}.
As we also discussed in Paper II, the variability in the line profile and flux of the 
v=1-0 excess emission component may be due to 
time variable circumplanetary disk accretion 
that results from either the predicted azimuthal variation in the 
accretion rate onto the circumplanetary disk (\citealt{KD06}) and/or 
non-steady accretion through the circumplanetary disk (\citealt{ML11a}). 

The interpretation that the excess v=1-0 CO emission 
arises from a circumplanetary disk raises a couple of questions.
One might wonder why the circumplanetary 
disk does not also produce observable CO hotband and OH 
emission. The lack of CO hotband emission is readily explained 
by the small emitting area (0.1\,AU$^2$) of the circumplanetary disk. 
Because the CO hotband emission is produced by 
UV fluorescence of disk gas,  the resulting infrared 
lines are very optically thin, and the emission must extend over a much
larger area to be observable. Our earlier detailed modeling of the 
hotband emission showed that the flux of the hotband lines 
could be explained by UV fluorescence of a thin layer of CO at 
the surface of the disk that extends from the disk wall  to 
$\sim 100$\,AU  \citep{Britt09}. 

The lack of a time variable OH excess component in the circumplanetary disk 
scenario is a more interesting question.  The circumplanetary disk may not 
produce much OH emission if its atmosphere has a high dust-to-gas ratio. 
Because an atmosphere with a high dust abundance is 
well shielded from UV radiation, OH production via water photodissociation is 
less efficient, leading to a low OH abundance.  
The outer disk, if it is less well shielded by dust (e.g., as a result of 
grain growth and settling), could then dominate 
the OH emission from the system. 

If the excess CO emission arises in a circumplanetary disk, 
observations of the kind presented here may lend insight into the 
nature of circumplanetary disks, which are thought to have 
complex dynamics. Hydrodynamical studies find 
that the accretion flow onto a planet is intrinsically 3-dimensional, 
with the circumplanetary disk having a large scale height ($H/R \sim 0.5$) and
sub-Keplerian velocities (e.g., \citealt{KK06, AB12}).  
Inflow toward the planet is found to occur 
primarily downward from high latitudes, with material flowing
outward in the midplane of the circumplanetary disk through most
(e.g.,\citealt{Tan12, Gressel13}) or only the outer
portion \citep{AB12} of the disk.  
These ideas could potentially be explored 
with observations similar to those presented here.  
Because of the complex
flow pattern, the expected line profile for CO emission from a
circumplanetary disk will depend on where warm, dense CO is 
located, i.e., on the thermal-chemical properties of the
region.

The limited constraints on the orbital radius of the CO excess 
emission component and the position of the disk wall (Section 4.1)
allow for the possibility that the excess emission arises
not from a discrete source (e.g., a circumplanetary disk), but 
instead from a localized region of the circumstellar 
disk, e.g., a CO bright spot on the disk wall.
The bright spot could result 
from a local temperature or density enhancement. 
If it results from a temperature enhancement, both the CO and OH
emission would be increased over that of the surrounding disk
unless, e.g., the CO bright spot is also a region of high dust 
abundance, in which case the OH emission from the region 
may be suppressed (as discussed above). 

Alternatively, because the CO v=1-0 emission from HD~100546 appears to be 
subthermally excited (Brittain et al.\ 2009), the CO emission may be 
particularly sensitive to density enhancements. In contrast, 
the ro-vibrational OH emission is more likely to be in LTE 
given its much lower critical density\footnote{\citet{Atahan06} 
find that the collisional deexcitation rate out 
of OH v=1 at 300~K is $\rm 2\times10^{-10}~cm^3~s^{-1}~mol^{-1}$, 
or three orders of magnitude larger than the rate for CO at 300~K 
\citep{Thi13}. Given the similarity of the transition probabilities for OH and CO, 
this implies that the critical density of OH is three 
orders of magnitude lower than that of CO at this temperature. 
Modeling of the excitation of OH requires determination of the 
collisional deexcitation rate at higher temperatures}  
and therefore insensitive to density. 
Detailed excitation modeling of the CO and OH emission from HD~100546 
could be used to explore this and other potential explanations. 
If either a temperature or density enhancement turns out to be a viable 
explanation for the bright spot, 
it would be interesting to understand the physical 
origin of such a persistent enhancement, as the CO excess emission 
has now been observed over a 10 year interval. 
Such enhancements might be induced by an orbiting giant planet 
(e.g., a local hot spot) or arise in structures associated with the planet 
(e.g., gap-crossing streams). 
 
Higher angular resolution observations (e.g., with MagAO, GPI, SPHERE) can 
explore these scenarios by looking for evidence for an orbiting companion, 
a circumplanetary disk, or an emission bright spot on the disk wall. 
Observations made before 2017 have the best chance to detect 
the proposed orbiting structure before it is hidden by the disk wall.
If the orbiting companion that we infer from our data is confirmed, 
along with that observed by Quanz et al. (2013) at larger orbital radii, we may be 
witnessing an example of multiple (perhaps even related, or sequential) planet 
formation in a circumstellar disk.

\acknowledgments
Based on observations collected at the European Organization for Astronomical Research in 
the Southern Hemisphere, Chile, under program number 090.C-0571(A). S.D.B.  acknowledges support for 
this work from the National Science Foundation under grant number  AST-0954811. Basic research in 
infrared astronomy at the Naval Research Laboratory is supported by 6.1 base funding. JN gratefully 
acknowledges support from the Institute for Theory and Computation at the Harvard-Smithsonian 
Center for Astrophysics. 

{\it Facilities:} \facility{VLT:Antu (CRIRES)}.

\begin{table*}
\begin{center}
\caption[HD~100546: Log of Observations]{Log of Observations}\label{tab:observations}
\begin{tabular}{lccccccc}
	\hline
	Setting & Spectral Grasp & Integration  & S/N & Seeing & Centroid rms & Airmass \\
	& cm$^{-1}$&  minutes  &  &  arcsec & mas & sec(z) \\
	\hline \hline
	L1 & 3135-3207 &  24 & 400  & 0.55 & \nodata & 1.72 \\	
	L2 & 3365-3448 &  35 & 230  & 0.73 & \nodata & 2.08 \\
	M1 & 1998-2053 &  64 & 175  & 0.52 & 0.86 & 1.53 \\
	\hline
\end{tabular}
\end{center}
\end{table*}%

%
\begin{table*}
\begin{center}
\caption[HD~100546: OH Lines]{Flux of OH Emission lines}\label{tab:ohlines}
\begin{tabular}{lccc}
	\hline
	Line & Wavenumber & EqW & F$\rm_{line}$   \\
	& cm$^{-1}$ & 10$^{-3}$cm$^{-1}$ & 10$^{-15}$ erg/s/cm$^2$ \\
	\hline \hline
P4.5(1+)	&	3407.99	&	4.56	$\pm$	0.46	&	5.04	$\pm$	0.50	\\
P4.5(1-)	&	3407.61	&	5.99	$\pm$	0.60	&	6.62	$\pm$	0.66	\\
P8.5(2-)	&	3194.28	&	2.97	$\pm$	0.30	&	3.27	$\pm$	0.33	\\
P8.5(2+)	&	3193.69	&	3.10	$\pm$	0.31	&	3.41	$\pm$	0.34	\\
P9.5(2-)	&	3146.18	&	1.55	$\pm$	0.15	&	1.70	$\pm$	0.17	\\
P9.5(2+)	&	3145.49	&	2.62	$\pm$	0.26	&	2.88	$\pm$	0.29	\\
P10.5(1+)	&	3142.06	&	1.81	$\pm$	0.54	&	1.98	$\pm$	0.59	\\
P10.5(1-)	&	3141.04	&	2.85	$\pm$	0.57	&	3.13	$\pm$	0.63	\\
	\hline

\end{tabular}
\end{center}
\end{table*}%

\begin{table*}
\begin{center}
\caption[HD~100546: Comparison of excess Equivalent Width of P26 line ]{Properties of Excess CO v=1--0 Emission}\label{tab:excess_eqw}
\begin{tabular}{p{2.35cm}p{1.5cm}p{1.5cm}p{1.1cm}p{1.2cm}p{1.2cm}p{1.3cm}p{1.5cm}p{1.5cm}}
	\hline
	Date& Scaled Equivalent Width$^1$ & Excess Equivalent Width & Doppler Shift of Excess & FWHM of Excess & Position Angle of Excess & Orbital Phase of Excess$^2$ & Red Offset & Blue Offset \\
	& 10$^{-2}$ cm$^{-1}$ & 10$^{-2}$ cm$^{-1}$ & km s$^{-1}$ & km s$^{-1}$ & & & mas & mas  \\
	\hline \hline
	2003 January 7	 	& 4.50$\pm$0.14 & \nodata  & \nodata  & \nodata & --40\degr & $0\degr$ & 12.7 $\pm$ 3.3 & 14.0 $\pm$ 3.3  \\	
	2006 January 14	 	& 5.69$\pm$0.59 & 1.19$\pm$0.61 & $+6\pm1$ & 6 & --5\degr & $47\degr \pm 10\degr $& 1.6 $\pm$ 4.3 & 17.8 $\pm$ 4.3 \\	
	2010 December 23	 	& 6.39$\pm$0.57 & 1.89$\pm$0.58 & $-1\pm1$ & 12 & 60\degr & $97\degr  \pm 7\degr  $&  5.7 $\pm$ 4.8 & 25.9 $\pm$ 4.8 \\
	2013 March 18		         &  5.89$\pm$0.20 & 1.12$\pm$0.15 & $-6\pm1$ & 6 & 105\degr & $133\degr  \pm 10\degr  $& 10.9 $\pm$ 0.9 & 35.8 $\pm$ 0.9      \\

\hline

\end{tabular}
\footnotetext{\footnotesize$^1$ The spectra were scaled such that the average hot band line profiles observed in 2006 and 2010 had the same equivalent widths as the average line profile observed in 2003.}
\footnotetext{\footnotesize$^2$ The phase is measured counter clockwise from the northwest end of the semimajor axis of the disk.}
\end{center}
\end{table*}%

\begin{figure*}
    \begin{center}
	\includegraphics[scale=.6]{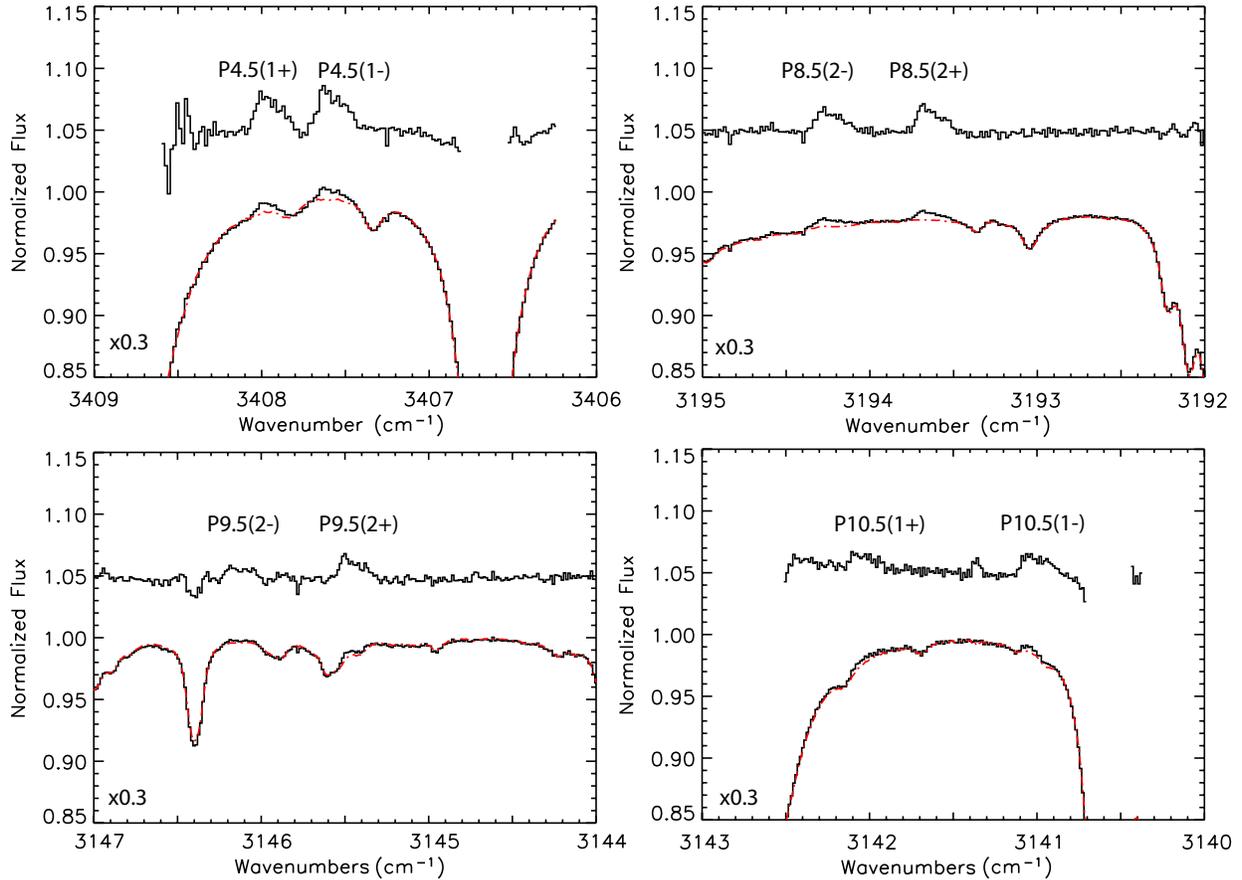}
	\caption{$L-$band Spectra of HD~100546. The regions of the $L-$band spectra containing v=1-0 OH emission lines are plotted. 
	The telluric standard (red, dot-dashed line) is plotted over the spectrum of HD~100546 (solid black line). The spectra are scaled by a factor of 
	0.3. The ratioed spectrum is plotted above the spectra and is offset +0.05 units. Regions where the atmospheric transmittance is less than 50\% are
	omitted. The asymmetric line profile is apparent in the individual spectra. }
	\label{fig:obs1}
   \end{center}
\end{figure*}

\begin{figure*}
    \begin{center}
	\includegraphics[scale=.6]{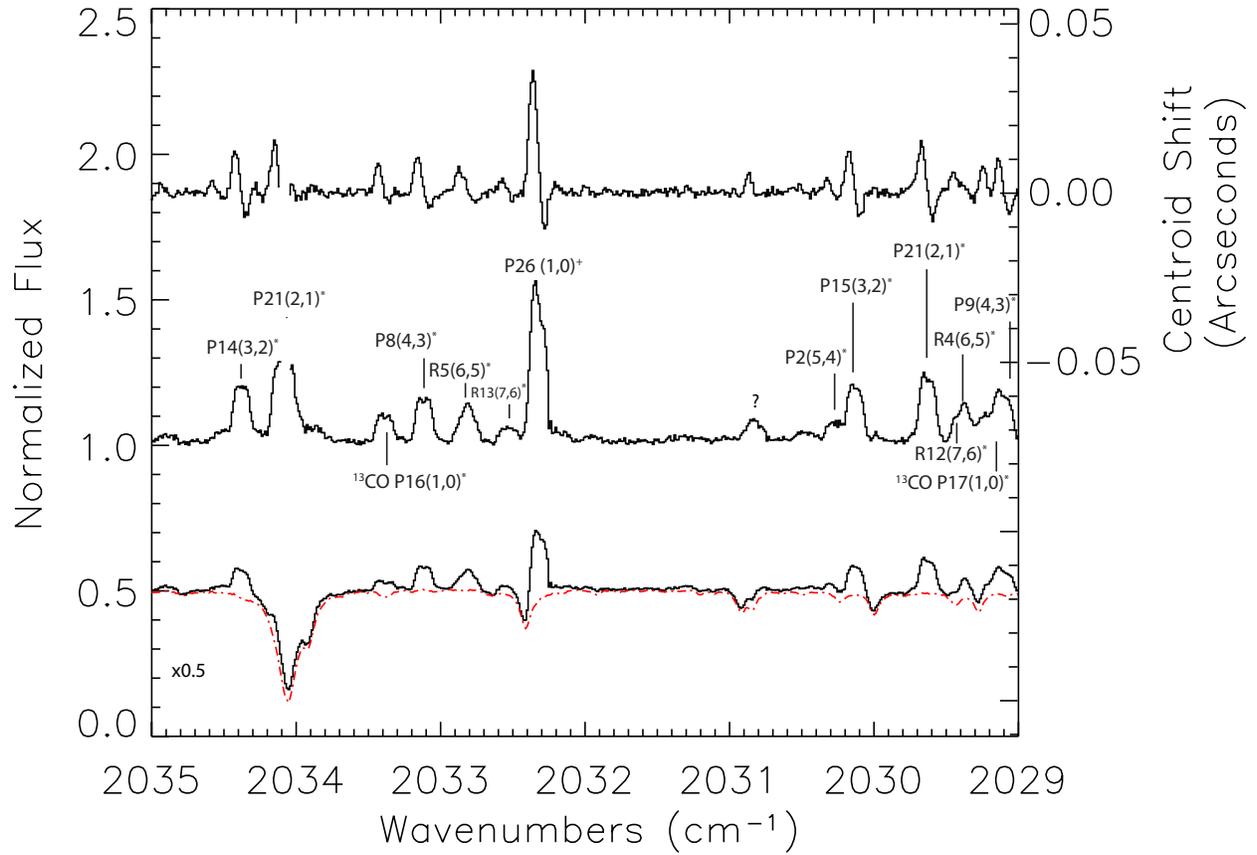}
	\caption{$M-$band spectrum of HD~100546. The portion of the $M-$band spectrum observed in previous 
epochs is plotted. The CO hotband and $^{13}$CO lines are indicated by asterisks; these lines are excited by UV 
fluorescence. The most prominent emission line in the spectrum is the v=1-0 P26 line near
2032.35 cm$^{-1}$ (labeled with a "+"),  which is excited by a
combination of UV fluorescence and collisions. The spectrum of
HD~100546 (solid black) and the telluric spectrum $\alpha$Cru
(dot-dashed red) are scaled by a factor of 0.5 and plotted below
the ratioed spectrum. Plotted above the ratioed spectrum is the
spectro-astrometric signal of the spectrum taken with the slit
rotated to a position angle 90$\degr$ E of N. The hotband lines are slightly asymmetric, and 
the v=1-0 P26 line shows a much more dramatic asymmetry. }

	\label{fig:obs2}
   \end{center}
\end{figure*}

\begin{figure*}
    \begin{center}
	\includegraphics[scale=.5]{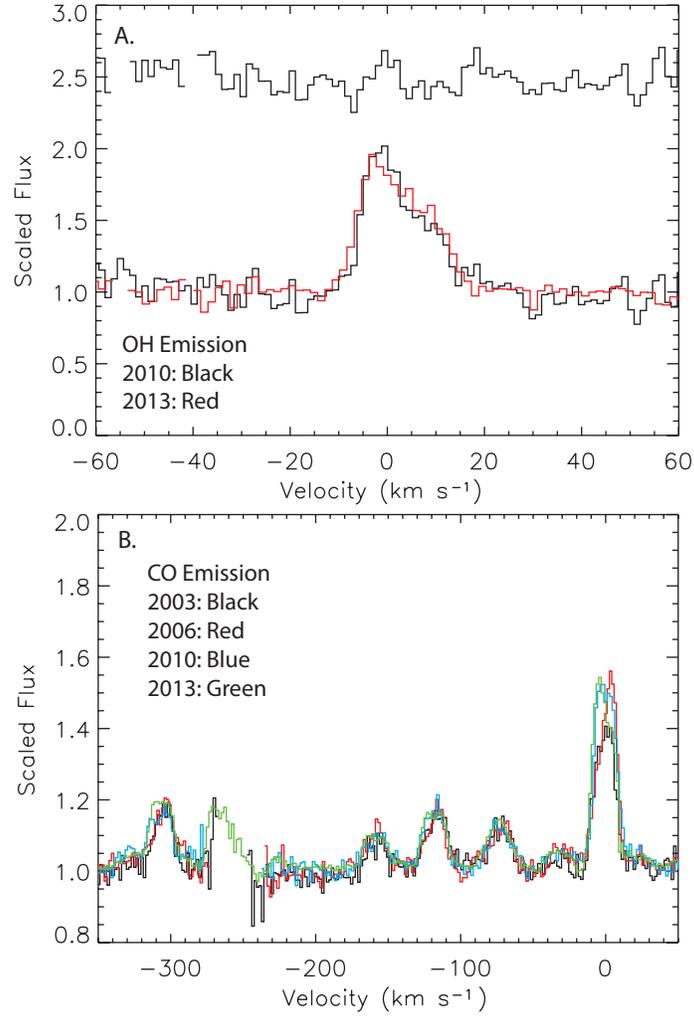}
	\caption{Multi-epoch observations of the of the OH (Panel A) and CO (Panel B) lines. In Panel A, the average of the OH emission lines observed in 2010 (black) and 2013 (red) are plotted over one another. Both lines have been scaled to a constant equivalent width.  The difference between these spectra is plotted above. While the equivalent width of the lines varied, the shape of the lines has not varied to within the signal to noise of our measurement. In Panel B we plot the overlapping region of the CO spectra observed over four epochs. The spectra have been scaled so that the equivalent width of the average of the hotband lines is constant. While the shape of the hotband lines has not changed over the four epochs spanning 2003-2013, the v=1-0 P26 line has varied. In 2006, the P26 line shows a red excess relative to the 2003 spectrum. In 2010, the excess shows a minimal Doppler shift ($-1\pm1$km s$^{-1}$) relative to 2003. In 2013, the P26 line shows a blue shifted excess.}
	\label{fig:obs3}
   \end{center}
\end{figure*}

\begin{figure*}
    \begin{center}
	\includegraphics[scale=.7]{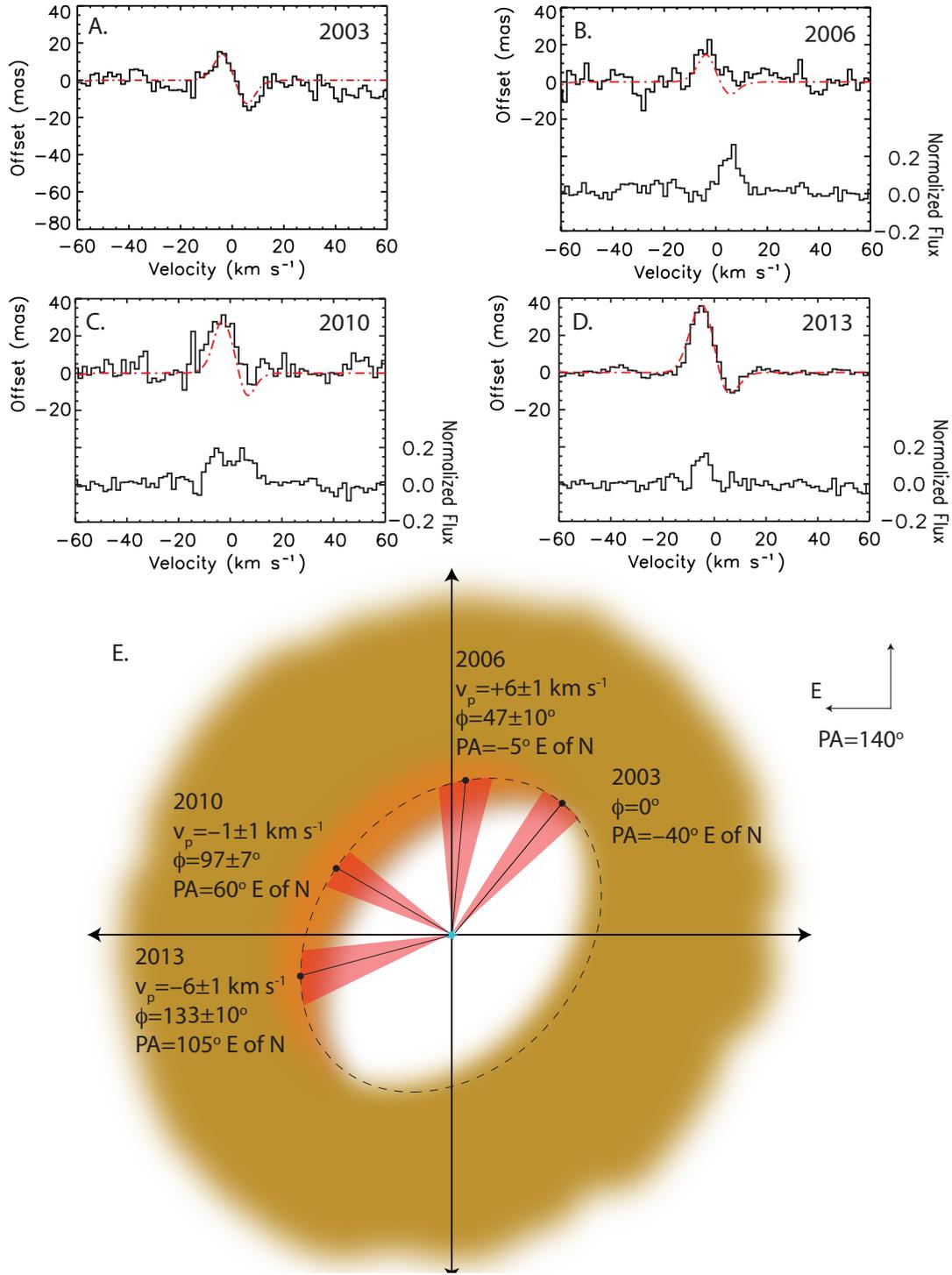}
	\caption{Spectroastrometric signal of the P26 line and 
schematic of the geometry of the system. In Panels A--D the
spectro-astrometric signal of the P26 line is plotted. The
excess flux of the P26 line is plotted below the
spectro-astrometric signal in Panels B--D. For each epoch
the spectro-astrometric signal is calculated from our
excitation model with the excess emission added for the
data acquired in 2006, 2010, and 2013 (red dot-dashed line).
In Panel E, a schematic of the disk and extra emission
source is presented. The orbit is represented by the black
dashed line. The disk wall of the disk is shaded 
orange. The location of the source of the emission excess is labeled with a black
dot, and the uncertainty in the phase of the orbit is
represented by the red sectors. We assume the 
excess CO emission is hidden by the near side of the circumstellar
disk in 2003. The phase of the orbit is calculated from the Doppler
shift of the excess emission assuming the disk is inclined
by 42$\degr$ and the orbital radius is 12.5~AU (just inside
the disk wall of the disk).  In 2006, the excess emission
pulled the center of light of the red side of the line
closer to the the center of the PSF. In 2010, the excess
emission pulled both sides of the line eastward along the
slit axis.  In 2013, the excess emission on the blue side
of the line pulled the spectro-astrometric signal eastward.  }
	\label{fig:obs5}
   \end{center}
\end{figure*}


\begin{figure*}
    \begin{center}
	\includegraphics[scale=.4]{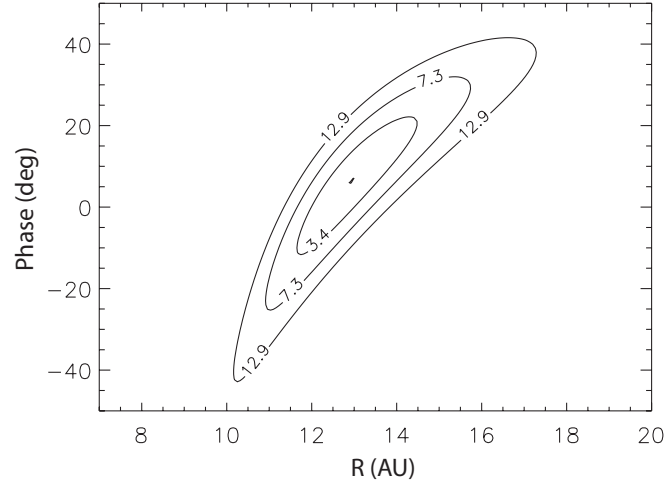}
	\caption{Constraint on the orbital radius $R$ and orbital phase in 2003 $\phi_0$ 
placed by fitting the projected velocities and relative phases in 2006, 2010, and 2013, 
assuming the excess CO emission is 
in a circular Keplerian orbit and the inclination is 42$\degr$. The 1$\sigma$, 2$\sigma$, and 3$\sigma$ 
	confidence intervals are plotted. We find that R$=12.9^{+1.5}_{-1.3}$
	~AU and $\phi=6\degr^{+15\degr}_{-20\degr}$.}
\label{fig:obs5}
   \end{center}
\end{figure*}

\end{document}